\documentclass[twocolumn]{aastex631}
\usepackage{amsmath}
\usepackage{graphicx}
\usepackage{hyperref}
\usepackage{booktabs}
\defcitealias{Cehula+24}{C24}

\newcommand{\be}{\begin{equation}}
\newcommand{\ee}{\end{equation}}

\shorttitle{evidence for \textit{r}-process nucleosynthesis in the SGR 1806-20 magnetar giant flare }
\shortauthors{A.~Patel et al.}

\begin{document}

\title{Direct evidence for \textit{r}-process nucleosynthesis in delayed MeV emission from the SGR 1806-20 magnetar giant flare}

\author[0009-0000-1335-4412]{Anirudh Patel}
\affil{Department of Physics and Columbia Astrophysics Laboratory, Columbia University, New York, NY 10027, USA}

\author[0000-0002-4670-7509]{Brian D.~Metzger}
\affil{Department of Physics and Columbia Astrophysics Laboratory, Columbia University, New York, NY 10027, USA}
\affil{Center for Computational Astrophysics, Flatiron Institute, 162 5th Ave, New York, NY 10010, USA} 

\author[0000-0002-4914-6479]{Jakub Cehula}
\affil{Institute of Theoretical Physics, Faculty of Mathematics and Physics, Charles University, V Hole\v{s}ovi\v{c}k\'{a}ch 2, Prague, 180 00, Czech Republic}

\author[0000-0002-2942-3379]{Eric Burns}
\affil{Department of Physics \& Astronomy, Louisiana State University, Baton Rouge, LA 70803, USA}

\author[0000-0003-1012-3031]{Jared A.~Goldberg}
\affil{Center for Computational Astrophysics, Flatiron Institute, 162 5th Ave, New York, NY 10010, USA} 

\author[0000-0002-1730-1016]{Todd A.~Thompson}
\affil{Department of Astronomy, Ohio State University, 140 West 18th Avenue, Columbus, OH 43210, USA}
\affil{Center for Cosmology \& Astro-Particle Physics, Ohio State University, 191 West Woodruff Ave., Columbus, OH 43210, USA}
\affil{Department of Physics, Ohio State University, 191 West Woodruff Ave., Columbus, OH 43210, USA}

\correspondingauthor{Anirudh Patel}
\email{anirudh.p@columbia.edu}

\begin{abstract} 

The origin of heavy elements synthesized through the rapid neutron capture process ($r$-process) has been an enduring mystery for over half a century. \citet{Cehula+24} recently showed that magnetar giant flares, among the brightest transients ever observed, can shock-heat and eject neutron star crustal material at high velocity, achieving the requisite conditions for an $r$-process. \citet{Patel+2025} confirmed an $r$-process in these ejecta using detailed nucleosynthesis calculations. Radioactive decay of the freshly synthesized nuclei releases a forest of gamma-ray lines, Doppler broadened by the high ejecta velocities $v \gtrsim 0.1c$ into a quasi-continuous spectrum peaking around 1 MeV. Here, we show that the predicted emission properties (light-curve, fluence, and spectrum) match a previously unexplained hard gamma-ray signal seen in the aftermath of the famous December 2004 giant flare from the magnetar SGR 1806-20. This MeV emission component, rising to peak around 10 minutes after the initial spike before decaying away over the next few hours, is direct observational evidence for the synthesis of $\sim 10^{-6}M_{\odot}$ of $r$-process elements. The discovery of magnetar giant flares as confirmed $r$-process sites, contributing at least $\sim 1$--$10\%$ of the total Galactic abundances, has implications for the Galactic chemical evolution, especially at the earliest epochs probed by low-metallicity stars. It also implicates magnetars as potentially dominant sources of heavy cosmic rays. Characterization of the $r$-process emission from giant flares by resolving decay line features offers a compelling science case for NASA's forthcoming COSI nuclear spectrometer, as well as next-generation MeV telescope missions.

\end{abstract}

%\keywords{}

\section{Introduction} \label{sec:intro}

Roughly half of the elements in our universe heavier than iron are synthesized through the rapid neutron capture process ($r$-process; \citealt{Burbidge+57,Cameron57}). Despite this recognition, identifying the astrophysical sites that give rise to the necessary conditions for an $r$-process has remained challenging.  Possibilities include neutron star mergers \citep{Lattimer&Schramm74,Symbalisty&Schramm82,Eichler+89}, proto-neutron star winds during core-collapse supernovae \citep{Meyer+92, Qian&Woosley96, Prasanna+2024}, and black hole accretion disk outflows in collapsars (e.g., \citealt{Siegel+19}), among other sources.  However, there has been little observational evidence for the $r$-process in action, with the notable exception of the kilonova signal following neutron star mergers \citep{Li&Paczynski98,Metzger+10,Tanvir+13,Berger+13,Levan+24}, particularly that which accompanied the LIGO merger GW170817 \citep{Coulter+2017,Kasen+17}.  Even in well-studied kilonovae, identification of individual atomic spectral features has proven challenging \citep{Hotokezaka+2022, Tarumi+23,Levan+24}, although some studies suggest a possible detection of a light $r$-process feature (e.g., \citealt{Watson+19}). 

While neutron star mergers are likely a major $r$-process source in our Galaxy, chemical evolution studies using stellar abundances and isotopic analyses of meteorites have been used to argue they are not the only one in operation \citep{Cote+19,Zevin+2019,vandeVoort+20,Tsujimoto+2021,Simon+2023}, particularly at low metallicities \citep{Qian&Wasserburg07, Sneden+08,Thielemann+2020,Ou+2024}. Mergers with short delay times $\lesssim 100$ Myr can enrich metal-poor stars with $r$-process elements (e.g., \citealt{Beniamini&Piran24}), in addition to other sources which directly track star formation.

Magnetars are the most highly magnetized neutron stars in our Galaxy, with surface dipole magnetic field strengths $B \sim 10^{14}$--$10^{15} \,\rm G$ \citep{Duncan&Thompson92,Kouveliotou+98}. They exhibit a range of transient X-ray and gamma-ray outbursts, powered by dissipation of their ultra-strong magnetic fields  \citep{Mereghetti+2015,Turolla+2015,Kaspi2017}. The most rare and energetic magnetar outbursts are the ``giant flares," which release $\sim 10^{44}\text{--}10^{46}\,\rm ergs$ in gamma-rays in under a second. Seven extragalactic giant flare candidates have been identified \citep{Burns+21,Beniamini+2024,Rodi+2024} and three have been observed within our Galaxy or the Large Magellanic Cloud: in 1979 \citep{Mazets+1979, Evans+1980}, 1998 \citep{Hurley+1999}, and 2004 \citep{Palmer+2005,Hurley+05}. All three events comprised a brief ($\lesssim 0.5$ s) gamma-ray burst followed by a $\sim$ minutes-long pulsating hard X-ray tail modulated by the magnetar spin period. The two Galactic giant flares were followed by a synchrotron radio afterglow lasting several months \citep{Frail+1999,Hurley+1999, Cameron+2005,Gaensler+2005}, which arose from shock interaction between the material ejected during the flare  and the surrounding medium. Modeling of the radio emission suggests baryon-loaded ejecta with a mass of $\sim 10^{24.5}$--$10^{27}$ g and velocities up to $v \sim 0.5 \text{--} 0.7c$ \citep{Gelfand+2005, Taylor+2005,Granot+2006}.

Motivated by these observations, \citet[hereafter \citetalias{Cehula+24}]{Cehula+24} developed a model for baryon ejection in magnetar giant flares, supported with hydrodynamical simulations. They postulate that the $e^{\pm}$ pair-photon fireball generated above the neutron star surface during the flare drives a shockwave into the crust, which they show can heat $\sim 10^{-8}\text{--}10^{-6}M_{\odot}$ of baryonic matter to sufficient energy to escape the gravitational potential well with a range of velocities $v \gtrsim 0.1c$. \citetalias{Cehula+24} further showed that the conditions in the expanding baryonic debris are promising for $r$-process nucleosynthesis. 

Although the outer neutron star crust ejected during a giant flare is not necessarily neutron-rich (electron fraction $Y_e \gtrsim 0.40$ for spherically symmetric ejection), the synthesis of heavy elements up to and beyond the 2nd $r$-process peak ($A \gtrsim 130$) is nevertheless possible via the ``$\alpha$--rich freeze-out'' mechanism. In this mechanism, the formation of seed nuclei onto which neutrons are captured is suppressed by the high entropy (low density) and fast expansion rate of the ejecta (e.g., \citealt{Hoffman+97}), thereby raising the neutron-to-seed ratio and enabling synthesis of heavier nuclei during the $r$-process. Using dynamical and thermodynamic ejecta properties motivated by \citetalias{Cehula+24}, \citet{Patel+2025} employed a nuclear reaction network to confirm the $r$-process in magnetar flare ejecta and quantify the synthesized abundances. 

Freshly synthesized $r$-process nuclei are radioactive, and they release energy continuously over their wide range of decay half-lives, from less than milliseconds to millions of years or longer, in the form of $\beta$--decay electrons, neutrinos, and nuclear gamma-ray lines \citep{Metzger+10,Hotokezaka+16}. Early after ejection from the neutron star, the expanding debris is opaque to gamma-rays, resulting in their energy being efficiently thermalized with the plasma. This radioactive heating, along with that from energetic electrons, powers a luminous optical/UV-wavelength transient lasting minutes after the flare, akin to a scaled-down kilonova; this {\it nova brevis} signal peaks a few minutes after the flare at a luminosity $\sim 10^{38}$--$10^{40}$ erg s$^{-1}$ and may be detectable with wide-field optical/UV telescopes for sources out to several Mpc (\citetalias{Cehula+24}, \citealt{Patel+2025}).  

After expanding and decompressing for several minutes, the ejecta will become transparent, enabling gamma-rays to escape unattenuated. Prospective observations of escaping gamma-rays would offer one of the cleanest and most direct probes of the $r$-process products \citep{Qian+98,Hotokezaka+16,Wu+2019,Li19,Korobkin+2020, Chen+2021, Terada+22,Chen+2024}. In particular, the gamma-ray spectrum can reveal line signatures of individual radioactive isotopes, provided they can be identified despite the Doppler broadening they experience from the high ejecta velocities $v \gtrsim 0.1c$. 

The remarkable finding presented here is that such a nuclear decay-line signal may already have been detected 20 years ago, in the aftermath of the 2004 giant flare from SGR 1806-20 \citep{Hurley+05,Palmer+2005}. Starting around $t \approx 400\,\rm s$ after the initial gamma-ray spike, and following the decay of the pulsating X-ray tail, the anticoincidence shield (ACS) on the {\it INTEGRAL} satellite observed a new emission component appear in the gamma-ray light-curve \citep{Mereghetti+05}. This component rose to a broad peak around $t \approx 600\text{--}800\,\rm s$, before decaying below the background level at $t \gtrsim 3000\text{--}8000\, \rm s$, roughly with flux $\propto t^{-\delta}$ with $\delta = 1.2 \pm 0.1$ (Appendix \ref{sec:app}).

\begin{figure*}
    \centering
    \includegraphics[width=1.\textwidth]{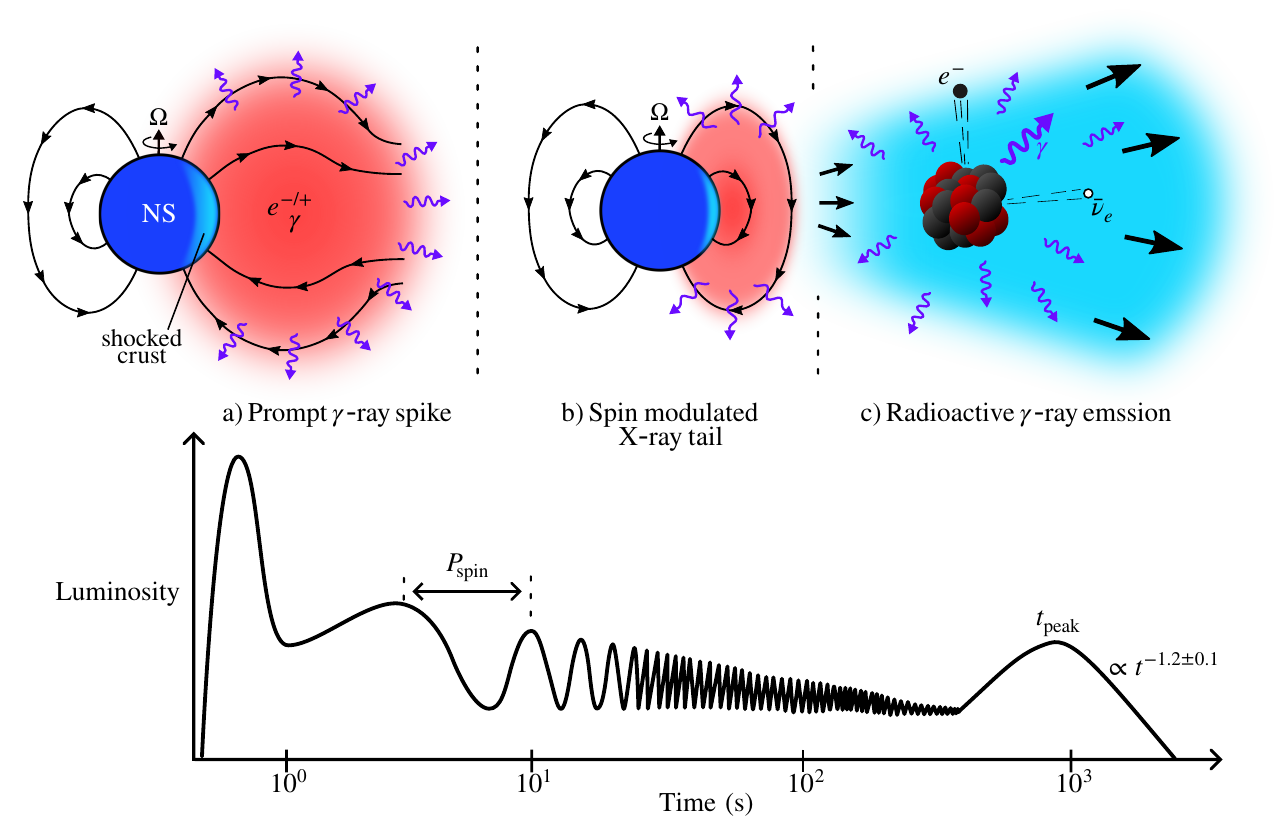}
    \caption{Schematic figure showing the three phases of high-energy emission following magnetar giant flares, as observed in the 2004 event from SGR 1806-20. After the prompt $\lesssim 1\,\rm s$ gamma-ray spike, and the minutes-long pulsating X-ray tail modulated on the neutron star rotation period, a third emission phase was observed in the form a smoothly evolving MeV component \citep{Mereghetti+05,Frederiks+07,Boggs+07}. This delayed MeV emission rose to a peak luminosity over $t\approx 600$--$800\, \rm s$, thereafter decaying smoothly until fading below the instrumental background a few hours later (see Appendix \ref{sec:app} and the bottom panel of Fig.~\ref{fig:lightcurve}). The mechanism of baryon ejection shown in Panel A is uncertain, but shock-heating of the neutron star crust by energy released during magnetic reconnection is one possibility (\citetalias{Cehula+24}) which would naturally lead to both a spin modulated X-ray tail and, as we argue here, delayed gamma-rays from the freshly synthesized radioactive $r$-process material.  We have illustrated the ejection as equatorial, but the direction of the mass ejection relative to the magnetic and spin axes is uncertain.}
    \label{fig:cartoon}
\end{figure*}

While the {\it INTEGRAL} ACS observations contain little spectral information, the delayed MeV component was also detected by {\it Konus-WIND} over the $t \sim 5\text{--}12\,\rm ks$ interval (the lack of earlier signal being a data limitation; \citealt{Frederiks+07}), who reported a fluence over this interval of $F_{\gamma} \approx 2\times 10^{-4}\,\rm erg \, cm^{-2}$ in the $80\text{--}750$ keV energy range. {\it RHESSI} also detected the late-time MeV component \citep{Boggs+07}, finding the spectrum around $t \sim 10^{3}\,\rm s$  could be modeled with thermal bremsstrahlung emission of temperature $k_{\rm B}T \approx 1.9\pm 0.7$ MeV with photons up to 2.5 MeV. From these datasets together, we estimate the total fluence out to 12 ks to be $F_{\gamma} \approx 9\times 10^{-4}\,\rm erg \, cm^{-2}$ (Appendix \ref{sec:app}). Quite distinct from the preceding trapped fireball X-ray phase, no pulsations were observed in the delayed MeV emission at the neutron star rotation period \citep{Boggs+07}.  Fig.~\ref{fig:cartoon} illustrates the three phases of the giant flare light-curve and the physical processes at work.  

In what follows, we show that all the key properties of the delayed MeV component from SGR 1806-20, namely its light-curve, fluence, and energy spectrum, are consistent with nuclear decay emission from freshly synthesized $r$-process ejecta. As we will describe, the discovery of magnetar giant flares as a confirmed $r$-process source has profound implications for the chemical evolution history of the universe, particularly those earliest stages following the first generations of stars.  Additional implications are discussed, including the origins of Galactic cosmic rays composed of $r$-process nuclei.

\section[Gamma-Ray Emission from Radioactive $r$-process Ejecta]%
{Gamma-Ray Emission from Radioactive \lowercase{$r$}-process Ejecta}
\label{sec:model}

We present calculations of the gamma-ray emission from freshly-synthesized $r$-process nuclei produced in magnetar giant flares, first using analytic arguments and then based on full nucleosynthesis calculations \citep{Patel+2025} with approximate radiation transport. Key results for the light curves and spectrum are presented in Figs.~\ref{fig:lightcurve}, \ref{fig:spectra}. 

\subsection{Light Curve and Fluence}
\label{sec:LC}

\begin{figure}
    \centering
    \includegraphics[width=0.5\textwidth]{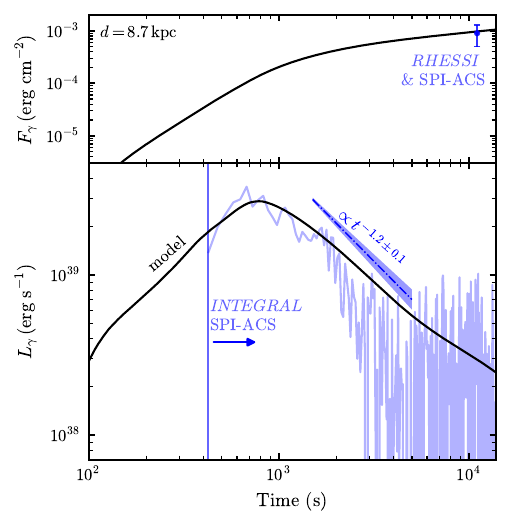}
\caption{Example $\sim$ MeV gamma-ray light-curve $L_{\gamma}(t)$ of $r$-process ejecta, calculated based on a multizone ejecta model from \citet{Patel+2025}, with $M_{\rm ej} \approx 1.2\times 10^{-6}M_{\odot}$, $M_{\rm r} \approx 7\times 10^{-7} M_{\odot}$, $\bar{v} = 0.15c$, $\beta = 6$, $f_{\Omega} = 3/4$.  Attenuation is treated approximately following \citet{Hotokezaka+16} for an assumed constant opacity $\kappa_{\gamma} = 0.1$ cm$^{2}$ g$^{-1}$. Shown with a purple line is the late-time MeV light-curve of the SGR 1806-20 giant flare detected by {\it INTEGRAL} ACS \citep{Mereghetti+05} over the time window $675$--$12000$ s, with the best fit decay rate $\propto t^{-1.2 \pm 0.1}$ in dark purple above the light-curve to guide the eye  (Fig.~\ref{fig:fit}; Appendix \ref{sec:app}). The top panel shows the cumulative fluence for an assumed source distance $d = 8.7$ kpc \citep{Bibby+08} in comparison to the estimated total fluence based on measurements from ACS and ${\it RHESSI}$ (\citealt{Boggs+07}; Appendix \ref{sec:app}).}
    \label{fig:lightcurve}
\end{figure}

The process by which baryons are ejected from the neutron star crust during a magnetar giant flare is complex and remains under active investigation (\citealt{Demidov2023}; \citetalias{Cehula+24}; see Fig.~\ref{fig:cartoon}). However, assuming mass ejection occurs during or promptly following the initial flare, the unbound ejecta will approach homologous expansion by the later times of interest here $t \gtrsim 100$\,s. We adopt a power-law density profile of the form 
\be
\rho(r,t) = \frac{\beta-3}{4\pi f_{\Omega}}\frac{M_{\rm ej}}{(\bar{v}t)^{3}}\left(\frac{v}{\bar{v}}\right)^{-\beta}, v > \bar{v},
\label{eq:rho}
\ee
where $\bar{v}$ is a characteristic minimum velocity, $M_{\rm ej}$ is the total ejecta mass, $f_{\Omega} \equiv \Delta \Omega/4\pi$ is the outflow covering fraction where $\Delta \Omega \le 4\pi$ is the solid-angle subtended by the ejecta, and the power-law index $\beta > 5$ depends on the details of the ejection process. In the numerical estimates to follow we adopt $\beta = 6$, though the qualitative results are not sensitive to this assumption (Appendix~\ref{app:param}). The fact that the pulsating X-ray tail emission was not blocked by the ejecta in either of the two Galactic giant flares supports a modest covering fraction $f_{\Omega} < 1$ \citep{Granot+2006}, consistent with the one-sided outflow inferred from VLBI imaging \citep{Taylor+2005}.

The optical depth seen by gamma-rays external to a given radius $r = vt$ is given by
\be
\tau_\gamma(r,t) = \int_{r}^{\infty}  \rho(r',t) \kappa_{\gamma}dr' = \frac{(\beta-3)}{(\beta-1)}\frac{M_{\rm ej}\kappa_{\gamma}(\bar{v}t)^{\beta-3}}{4\pi f_{\Omega} r^{\beta-1}},
\label{eq:tau}
\ee
such that the gamma-ray photosphere ($\tau_{\gamma} = 1$) is located at
\be
r_{\rm ph}(t) = \left(\frac{\beta-3}{\beta-1}\frac{M_{\rm ej}\kappa_{\gamma}}{4\pi f_{\Omega}}(\bar{v}t)^{\beta-3}\right)^{\frac{1}{\beta-1}} \propto t^{\frac{\beta-3}{\beta-1}},
\ee
corresponding to a velocity layer $v_{\rm ph}(t) = r_{\rm ph}/t \propto t^{-2/(\beta-1)}$ receding towards the center of ejecta with time. The bulk of the ejecta thus becomes optically-thin $(v_{\rm ph} \approx \bar{v}$) on the timescale:
\begin{eqnarray}
&& t_{\rm peak} \approx \left(\frac{\beta-3}{\beta-1}\frac{M_{\rm ej}\kappa_{\gamma}}{4\pi f_{\Omega}\bar{v}^{2}}\right)^{1/2} \nonumber \\
&\approx& 10^{3}\,{\rm s}\,\left(\frac{M_{\rm ej}/f_{\Omega}}{10^{-6}M_{\odot}}\right)^{1/2}\left(\frac{\kappa_{\gamma}}{0.1\,\rm cm^{2}\,g^{-1}}\right)^{1/2}\left(\frac{\bar{v}}{0.1c}\right)^{-1}.
\label{eq:tthin}
\end{eqnarray}
Gamma-rays lose energy to the plasma primarily through photoionization (bound-free) and (inelastic) Compton scattering, with a total opacity $\kappa_{\gamma} \approx 0.1\,\rm cm^{2}\, g^{-1}$ at photon energies $\approx 0.1$--$1$ MeV of interest (e.g., \citealt{Hotokezaka+16,Barnes+16}).

The specific radioactive energy generation-rate of $r$-process material can approximately be written \citep{Metzger+10,Roberts+11}
\be \label{eq:analytic_qdot}
\dot{q}_{\rm r}(t) \approx 5\times 10^{12}\left(\frac{t}{10^{3}\,{\rm s}}\right)^{-\alpha}\,{\rm erg\,s^{-1}\,g^{-1}},
\ee
where $\alpha \approx 1.1-1.4$ and the heating normalization depends on the time-frame under consideration and the synthesized abundances (e.g., \citealt{Barnes+21}).  We take $\alpha = 1.2$ and normalize $\dot{q}_{\rm r}$ based on our nucleosynthesis calculations (Sec.~\ref{sec:spectrum}). For a total mass $M_{\rm r}$ of $r$-process nuclei, the emitted gamma-ray luminosity can be written
\begin{eqnarray} \label{eq:analytic_Qgamma}
&& \dot{Q}_{\gamma}(t) = M_{\rm r}\epsilon_{\gamma}\dot{q}_{\rm r}(t) \nonumber \\
&\approx& 4\times 10^{39}\,{\rm erg\,s^{-1}}\left(\frac{\epsilon_{\gamma}}{0.4}\right)\left(\frac{M_{\rm r}}{10^{-6}M_{\odot}}\right)\left(\frac{t}{10^{3}\,{\rm s}}\right)^{-1.2}, 
\end{eqnarray}
where $\epsilon_{\gamma} = \dot{q}_\gamma/\dot{q}_{\rm r}\approx 0.3$--$0.4$ is the fraction of the decay-energy emitted as gamma-rays (\citealt{Barnes+16}, and verified with our numerical calculations; Appendix~\ref{sec:partition}). In the $\alpha$--rich freeze-out mechanism for enabling the $r$-process, a significant fraction of the ejecta mass resides in $\alpha$--particles and light seed nuclei, such that typically $M_{\rm r} \approx  0.3\text{--}0.8 M_{\rm ej}$ \citep{Patel+2025}.

Insofar that inelastic Compton scattering of $\sim 1$ MeV photons can be treated as an effective absorptive opacity when $\tau_{\gamma} > 1$ (e.g., \citealt{Metzger+10}), the bulk of the gamma-rays are trapped until $\tau_{\gamma} \sim 1$ at $t \approx t_{\rm peak}$. The total escaping gamma-ray energy in the time interval $t \in [t_{\rm peak},10t_{\rm peak}]$ is therefore approximately:
\begin{eqnarray}
&& E_{\gamma} = \int_{\rm t_{\rm peak}}^{10t_{\rm peak}} \dot{Q}_{\gamma}dt \nonumber \\
&\approx& 7\times 10^{42}{\rm erg}\left(\frac{\epsilon_{\gamma}}{0.4}\right)\left(\frac{M_{\rm r}}{10^{-6}M_{\odot}}\right)\left(\frac{t_{\rm peak}}{10^{3}\,{\rm s}\,}\right)^{-0.2},
\end{eqnarray}
resulting in a total gamma-ray fluence at Earth,
\begin{eqnarray}
&& F_{\gamma} = \frac{E_{\gamma}}{4\pi d^{2}} \approx 8\times 10^{-4}\,{\rm erg\,cm^{-2}}\,\times \nonumber \\
&& \left(\frac{M_{\rm r}}{10^{-6}M_{\odot}}\right)\left(\frac{t_{\rm peak}}{10^{3}\,{\rm s}}\right)^{-0.2}\left(\frac{d}{8.7\,{\rm kpc}}\right)^{-2}.
\label{eq:Fgamma}
\end{eqnarray}
where $d$ is the source distance of SGR 1806-20, normalized to the value $8.7\pm 1.5\,\rm kpc$ from \citet{Bibby+08}.

For $M_{\rm ej} \approx M_{\rm r} \approx 10^{-6}M_{\odot}$, $\bar{v}\approx 0.1\text{--}0.2c$ and $\kappa_{\gamma} \approx 0.1\,\rm cm^{2} g^{-1}$, we thus predict a light-curve peak time $\sim t_{\rm peak}$ around 1000\,s (Eq.~\eqref{eq:tthin}) with a total gamma-ray fluence $F_{\gamma} \approx 8\times 10^{-4}$ erg cm$^{-2}$. Both are in broad agreement with those measured for the delayed MeV emission from SGR 1806-20 (\citealt{Mereghetti+05,Frederiks+07, Boggs+07}; Appendix \ref{sec:app}).  The total ejecta mass falls within the range $\sim 10^{24.5}$--$10^{27}$ g inferred from the radio afterglow \citep{Gelfand+2005}, while the implied kinetic energy, $E_{\rm KE} \approx M_{\rm ej}\bar{v}^{2} \sim 10^{46}\,\rm erg$, is comparable to the isotropic energy of the prompt gamma-ray spike \citep{Palmer+2005,Hurley+05}.

Fig.~\ref{fig:lightcurve} shows an example gamma-ray light-curve calculation, in which $\dot{Q}_\gamma(t)$ is determined numerically by combining nucleosynthesis calculations \citep{Patel+2025} with gamma radiation data for individual nuclei, as described in the next section (Eq.~\eqref{eq:gamma}). The model shown assumes $M_{\rm ej} \approx 1.2 \times 10^{-6} M_{\odot}$, $M_{\rm r} \approx 7\times 10^{-7} M_{\odot}$, $\beta = 6$, $\bar{v} = 0.15c$, $f_{\Omega} = 0.75$. We assume an (initial) electron fraction $Y_e = 0.40$, matching that encountered in the neutron star crust at this mass depth below the surface (\citetalias{Cehula+24}).
To estimate the gamma-ray luminosity in the optically-thick rising phase of the light-curve, we employ the approximate gamma-ray thermalization treatment of \citet{Hotokezaka+16}. The attenuation experienced by each velocity layer of the ejecta is accounted for independently based on the time-dependent optical depth ahead of it $\tau_\gamma(v, t)$ (Eq.~\eqref{eq:tau}) assuming a constant opacity $\kappa_{\gamma} = 0.1\,\rm cm^{2} g^{-1}$ appropriate to $\sim 1$\,MeV photons. 

Shown for comparison are the peak timescale and time-averaged power-law decay $\propto t^{-\delta}$, $\delta = 1.2 \pm 0.1$ found by our analysis of the decay phase of the SPI-ACS light-curve (Appendix \ref{sec:app}).  The top panel of Fig.~\ref{fig:lightcurve} shows the cumulative gamma-ray fluence, in comparison to the total $\approx 9\times 10^{-4}$ erg cm$^{-2}$ that we estimate for the SGR 1806-20 delayed emission component out to $12\,\rm ks$ (Appendix \ref{sec:app}). The presented model therefore aptly explains all the basic features of the SGR 1806-20 delayed MeV emission component.  

Figure \ref{fig:params} (Appendix~\ref{app:param}) shows how the predicted light curve changes if we vary the ejecta mass, velocity dependent mass distribution $\beta$, characteristic velocity $\bar{v}$, or assumed electron fraction about the fiducial model assumptions. As expected from the analytic estimates above (Eqs.~~\eqref{eq:tthin}, \eqref{eq:Fgamma}), a larger ejecta mass increases the overall peak timescale and gamma-ray fluence, while a larger velocity decreases the peak timescale but only increases the fluence moderately.  A larger $Y_e$ produces a lower mass fraction of $r$-process elements, decreasing the radioactive heating efficiency somewhat, and slightly reducing the emitted fluence.  These dependencies illustrate the main degeneracies that exist in fitting the observed signal and instill confidence in our finding that $\sim 10^{-6}M_{\odot}$ of $r$-process production occurred from the SGR 1806-20 giant flare. 

\subsection{Energy Spectrum}
\label{sec:spectrum}

\citet{Patel+2025} employ the {\it SkyNet} nuclear reaction network \citep{Lippuner&Roberts17} to calculate the unique nucleosynthesis history and radioactive heating rate in each velocity layer of the ejecta, consolidating these layer-by-layer calculations to predict the total nucleosynthesis yield and {\it nova brevis} light-curve for models spanning a range of ejecta masses $\sim 10^{-8} \text{--} 10^{-6}M_\odot$. Here we extend their calculations to predict gamma-ray emission and create synthetic spectra at distinct epochs in the decay evolution corresponding to the observed delayed MeV emission component. We use fiducial model parameters as introduced in the previous section ($M_{\rm ej} \approx 1.2\times10^{-6}M_\odot$, $\bar{v} = 0.15c$, $\beta = 6$), which provide an adequate fit to the gamma-ray light-curve (Fig.~\ref{fig:lightcurve}).  

We consider all nuclear species that contribute at least $1\%$ to the total radiation on the timescales of interest, $10\,{\rm s}\lesssim t\, \lesssim 14000\,\rm s$, by which point the $r$-process has ceased and the newly synthesized nuclei are decaying to stability, predominantly through $\beta^-$--decay.\footnote{Other decay modes, such as $\alpha$-emission and fission fragmentation, can be omitted since they contribute negligibly to the decay power in ejecta for which large quantities of the heaviest nuclei $A\gtrsim 190$ are not produced \citep{Barnes+16}.} For all nuclei of a species $i$, we assume the effective decay mode $i \rightarrow f +e^- + \bar{\nu}_e + \gamma$ in the quoted time range, representing a decay to an excited state of the daughter species $f$, followed by a prompt cascade to the ground state typically resulting in $\gamma$ emission.

The specific energy released per second as gamma radiation is then,
% \be
% % = \sum_i \dot{q}_{\gamma, i} (t)
% \dot{q}_\gamma (t)\, [{\rm erg\,s^{-1}\,g^{-1}}] = N_{\rm A} \sum_i Y_i(t) \lambda_i b_i\sum_j I_{ij} \varepsilon_{ij},
% \label{eq:gamma}
% \ee
\be
\dot{q}_\gamma (t) = N_{\rm A} \sum_i Y_i(t) \lambda_i b_i\sum_j I_{ij} \varepsilon_{ij},
\label{eq:gamma}
\ee
where $N_{\rm A}$ is Avogadro's (baryon) number, and $Y_i(t)$, $\lambda_i$, and $b_i$ are the time-dependent abundance per baryon, $\beta^-$--decay rate, and photon branching ratio (i.e., the fraction of $\beta^-$--decay energy emitted as photons) for species $i$, respectively. $I_{ij}$ is the relative intensity (per decay) of a photon with energy $\varepsilon_{ij}$, which depends on the nuclear level structure of species $i$. Weak reaction rates are taken from the JINA REACLIB database \citep{cyburt:10} and all other nuclear decay data are taken from the Evaluated Nuclear Structure Data Files.\footnote{\url{https://www.nndc.bnl.gov/ensdf/}. This data acquisition made extensive use of the pyNE software package.} 

The gamma-ray energy generation rate (Eq.~\eqref{eq:gamma}) is evaluated for each layer with its unique nucleosynthesis evolution, and ultimately combined to determine the fraction of $r$-process decay energy emitted as gamma radiation $\epsilon_\gamma(t)$ (Fig.~\ref{fig:epsilon}, Appendix~\ref{sec:partition}), the total gamma-ray luminosity $\dot{Q}_\gamma(t)$, and fluence over time $F_\gamma(t)$ for the total ejecta, akin to the same quantities estimated analytically in the previous section. These results for the fiducial model (with $\dot{Q}_\gamma$ attenuated to produce the final light-curve $L_\gamma$) are shown in Fig.~\ref{fig:lightcurve}.

We generate the spectra by logarithmically dividing energy bins between $1$--$10^4 \,\rm keV$ and counting the photon rate in each bin (i.e., in the decay of a single nucleus of species $i$, the $j$'th photon contributes $b_iI_{ij}$ counts to the bin corresponding to its energy $\varepsilon_{ij}$) at a given time. We account for Doppler broadening by spreading the spectral lines over a Gaussian of full-width half-maximum of $2\sqrt{{\rm ln}2}\, v/c$, where $v$ is the expansion velocity of the layer \citep{Hotokezaka+16}. 

\begin{figure}
    \centering
    \includegraphics[width=0.5\textwidth]{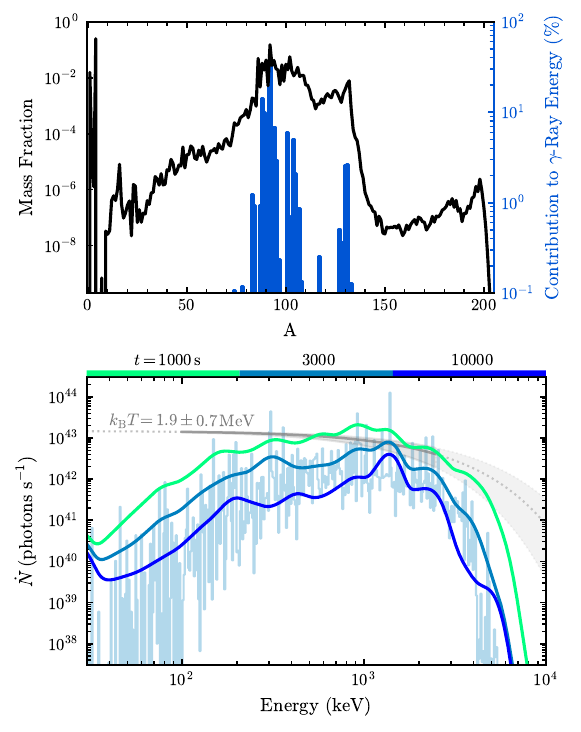}
\caption{{\it Top panel:} Mass fraction of synthesized nuclei as a function of atomic mass $A$ from our fiducial model which reproduces the late-time gamma-ray emission light-curve from SGR 1806-20 (Fig.~\ref{fig:lightcurve}). Vertical blue bars (right axis) show the individual percentage contributions of radioactive nuclei to the total gamma-ray energy in the time interval $t = 10^{3}$--$10^{4}\,$s. {\it Bottom panel:} Synthetic gamma-ray spectra from the fiducial model at three snapshots, $t = 1000,\, 3000,\, 12000\,$s, accounting for Doppler broadening due to the ejecta expansion but excluding extinction effects. For the spectrum at $t = 3000\,$s we also show with lighter blue lines the intrinsic (i.e., non-broadened) decay lines which contribute to the total spectrum. A dark gray band shows the bremsstrahlung spectral fit made by \citet{Boggs+07} in the energy range $\approx 0.1$--$2.5$ MeV, while a light gray band shows the same fit outside of the measured energy range.}
    \label{fig:spectra}
\end{figure}

The synthetic spectra for three different snapshots in time $t >t_{\rm peak}$ are shown in the bottom panel of Fig.~\ref{fig:spectra}. We include the $k_{\rm B}T = 1.9 \pm 0.7$ thermal bremsstrahlung fit to the {\it RHESSI} data reported in the $80$--$2500\,\rm keV$ range from \citet{Boggs+07}, showing qualitative agreement with our results. Although the photon flux observed at $\sim100~\rm keV$ slightly exceeds that in our predictions, the {\it RHESSI} spectrum is an empirical fit derived from a forward-folding approach, which was not informed by the model considered here, and the total errors are consistent with our model. Furthermore, Compton scattering at energies $\sim 1~\rm MeV$, would act to soften the emergent spectrum versus the intrinsic (non-attenuated) spectrum shown here. Since photons were not observed above $2600~\rm keV$, the reported fit is also uncertain at higher energies.

In the top panel of Fig.~\ref{fig:spectra}, we show the synthesized abundance distribution along with the fractional contribution to the total $\gamma$-ray energy by mass number (in the time interval $10^{3}$--$10^{4}\,\rm s$). While a robust $r$-process is realized up to the second peak ($A\sim 130$), and a moderate third peak ($A\sim 190$) yield is achieved, the MeV emission is dominantly powered by first peak nuclei ($A\sim 90$). Isotopes predicted to contribute substantially to the emission and their prominent decay lines producing discernible ``bumps" in the broadened spectra are listed in Table~\ref{table:isotopes}.

The detailed features of the gamma-ray spectra in Fig.~\ref{fig:spectra} are sensitive to the exact abundance yield and thus the ejecta properties. Future improvements on the mass ejection model of \citetalias{Cehula+24} including multidimensional and magnetohydrodynamic effects will likely inform different ejecta properties (e.g., $Y_e$, which is sensitive to the more complex mass excavation geometries present in multiple dimensions). However, we note the spectral peak near $\sim 1~\rm MeV$ and sharp cutoff above this generically arise from a large statistical ensemble of $r$-process nuclei, reflecting the characteristic gamma transitions energetically favored for $\beta$-decay products. Indeed, our predicted spectra are qualitatively similar to those of kilonova outflows, despite their different abundance yields (e.g., \citealt{Hotokezaka+16, Korobkin+2020}). Our identification of the delayed gamma-ray signal from SGR 1806-20 as radioactive decay emission is thus likely to remain robust.

\begin{table}[h!] 
\centering
\caption{Prominent Spectral Lines ($t \approx 10^{3}\text{--}10^{4}\,$s)}
\begin{tabular}{ccccc}
\toprule
Isotope & Daughter & $t_{1/2}$ (s) & Energy (keV) & Intensity\footnote{Absolute intensity, $b_iI_{ij}$ (Eq.~\eqref{eq:gamma}), per 100 counts.} \\
\midrule
$^{88}\text{Kr}$ &$^{88}\text{Rb}$ & $10170$ & $27.5$ & $1.9$ \\
    &&& $196.3$ & $26.0$ \\
    \midrule
$^{89}\text{Rb}$ & $^{89}\text{Sr}$& $919$ & $657.8$ & $10.8$ \\
    &&& $1031.9$ & $62.9$ \\
    &&& $1248.1$ & $45.9$ \\
    &&& $2195.9$ & $14.5$ \\
    &&& $2570.2$ & $10.7$ \\
$^{90}\text{Rb}$ & $^{90}\text{Sr}$& $158$ & $831.7$ & $39.9$ \\
    &&& $3383.2$ & $6.7$ \\
    &&& $3534.2$ & $4.0$ \\
    &&& $4135.5$ & $6.7$ \\
    &&& $4365.9$ & $8.0$ \\
    &&& $4646.5$ & $2.3$ \\
    &&& $5187.4$ & $1.2$ \\
    \midrule
$^{92}\text{Sr}$ & $^{92}\text{Y}$& $9396$ & $1383.9$ & $90.0$ \\
$^{93}\text{Sr}$ & $^{93}\text{Y}$& $445$ & $168.5$ & $18.4$ \\
    &&& $260.1$ & $7.4$ \\
    \midrule
$^{94}\text{Y}$ & $^{94}\text{Zr}$& $1122$ & $550.9$ & $4.9$ \\
    &&& $918.7$ & $56.0$ \\
    &&& $1138.9$ & $6.0$ \\
    \midrule
$^{101}\text{Mo}$ & $^{101}\text{Tc}$& $877$ & $9.3$ & $2.1$ \\
    &&& $80.9$ & $3.7$ \\
    &&& $191.9$ & $18.2$ \\
    \midrule
$^{101}\text{Tc}$ & $^{101}\text{Ru}$& $852$ & $306.8$ & $88.7$ \\
$^{104}\text{Tc}$ & $^{104}\text{Ru}$& $1098$ & $358.0$ & $89.0$ \\
    \midrule
$^{128}\text{Sn}$ & $^{128}\text{Sb}$& $3544$ & $32.1$ & $4.0$ \\
    &&& $45.7$ & $13.0$ \\
    &&& $75.1$ & $27.7$ \\
\bottomrule
\end{tabular}
    \label{table:isotopes}
\end{table}

\section{Discussion}

We have shown that the delayed gamma-ray emission component from the SGR 1806-20 magnetar giant flare is naturally understood as arising from the synthesis of $\sim 10^{-6}M_{\odot}$ of $r$-process material. We now describe some major implications of this finding and address alternative interpretations.

\subsection{Implications for Galactic Chemical Evolution}
\label{sec:chem_evolution}

The operation of magnetar giant flares as sources of $r$-process elements in our Galaxy has key implications for its chemical evolution. The Galactic rate of giant flares of energy $E_{\rm GF} \sim 10^{44}$--$10^{46}$ erg was estimated as (\citealt{Burns+21}) 
\be
\mathcal{R} \approx 8\times 10^{-3}{\rm yr^{-1}}\left(\frac{E_{\rm GF}}{10^{46}\rm erg}\right)^{-0.75},
\label{eq:R}
\ee
where we have given a median estimate, for which the 90\% uncertainty range for $E_{\rm GF} = 10^{46}$ erg is $(0.1-1.6)\times 10^{-2}$ yr$^{-1}$ (see also \citealt{Beniamini2019}).
If the $r$-process ejecta mass of a giant flare, $M_{\rm r}$, scales linearly (or super-linearly) with $E_{\rm GF}$ (e.g., \citetalias{Cehula+24}), then the total $r$-process yield $\propto \mathcal{R}M_{\rm r} \propto E_{\rm GF}^{0.25}$ over time would be dominated by the rarest, most energetic events.  In particular, based on the rate of powerful SGR 1806-20-like flares ($E_{\rm GF} \approx 10^{46}$ erg) from Eq.~\eqref{eq:R}, and the $r$-process yield $M_{\rm r} \approx 7\times 10^{-7}M_{\odot}$ inferred based on the radioactive gamma-ray signal (Fig.~\ref{fig:lightcurve}), we estimate a minimum present-day $r$-process production rate from Galactic magnetars of $\dot{M}_{\rm r} \sim \mathcal{R}M_{\rm r} \sim (0.7-11)\times 10^{-9}M_{\odot}$ yr$^{-1}$.  Over the history of the Galaxy, the time-averaged star-formation rate is $\approx 4$--$6$ times higher than the current rate (e.g., \citealt{Madau&Dickinson14}). Assuming that magnetar birth traces star-formation, we obtain a time-averaged $r$-process production rate $\langle \dot{M}_{\rm r} \rangle \approx 5 \dot{M}_{\rm r} \approx (0.4-6)\times 10^{-8}M_{\odot}$ yr$^{-1}$. This corresponds to $\sim 1-10\%$ of the total required $r$-process production rate in our Galaxy of $\langle \dot{M}_{\rm r,tot}\rangle \approx 5\times 10^{-7}M_{\odot}$ yr$^{-1}$, considering elements with atomic mass number $A \ge 90$ \citep{Hotokezaka+2018}. The implied fractional contribution would be higher if all flares (irrespective of energy) ejected a similar $r$-process mass as SGR 1806-20.

Although the uncertainties are large, magnetar flares are likely subdominant contributors to the Galactic $r$-process budget, compared to rarer but more prodigious sources such as neutron star mergers. However, giant flares are distinguished by operating with a short delay relative to star-formation: magnetars are among the youngest compact objects in the Galaxy, being frequently found still inside the remnants of their birthing supernovae (e.g., \citealt{Gaensler04}). Magnetars born of the first generation of stars thus offer a promising mechanism to enrich the most metal-poor stars in our Galactic halo.

Depending on their magnetic fields at birth $B \sim 10^{15}$--$10^{16}$ G and associated energy reservoir, a single magnetar can over its lifetime experience $\sim 10$--$10^{3}$ giant flares similar in strength to the 2004 flare from SGR 1806-20 (e.g., \citealt{Beniamini+20}). The total $r$-process production over a magnetar's $\sim 10^{4}$ yr active lifetime could thus be as great as $\sim 10^{-5}$--$10^{-3}M_{\odot}$; for context, this is comparable to the (potential) $r$-process yields of proto-neutron star winds (e.g., \citealt{Thompson+01}) but far lower than the typical yields of neutron star mergers or other rare events such as collapsars. Frequent, low-yield $r$-process sources operating in the early history of the Galaxy would help reconcile the otherwise puzzling observation that the stellar abundances of iron and $r$-process nuclei are observed to already begin correlating with each other at low metallicities [Fe/H] $\sim -3$ (e.g., \citealt{Qian&Wasserburg07}).

The synthesized abundance distribution of our fiducial model exhibits a pronounced peak around $A \sim 90$ (Fig.~\ref{fig:spectra}), corresponding to those nuclei (particularly Sr, Y, Zr) associated in the literature with being formed in the so-called ``light-element primary process'' (LEPP; \citealt{Arcones&Montes11}) and which have been previously attributed to proto-neutron star winds (e.g., \citealt{Witti+94}).  Our findings reveal that magnetar giant flares could also compete with supernovae as sources of LEPP nuclei, particularly if the electron fraction of proto-neutron star winds is too high $Y_{e}>0.5$ to enable their formation.

\subsection{Observations of Future Giant Flares }

To date, the 2004 event from SGR 1806-20 represents the only opportunity to detect the $r$-process transient from a magnetar giant flare.  {\it Konus-WIND} also took observations of the SGR 1900 giant flare in 1998, but the high radiation background due to enhanced solar activity would have precluded a detection of a similarly luminous delayed MeV component \citep{Frederiks+07}.  The duration and fluence of the SGR 1806-20 late-time emission are similar to those of the brightest long GRBs observed by {\it Swift BAT} \citep{Lien+16} and {\it Fermi GBM} \citep{Poolakkil+21}, a population which extends down to $F_{\gamma} \sim 10^{-6}$ erg cm$^{-2}$. An $r$-process gamma-ray signal similar to that in SGR 1806-20 could therefore be detected by these wide-field monitors out to a distance roughly 10 times further, i.e. to $\sim 100$ kpc. While this range encompasses the Magellanic Clouds, the $r$-process gamma-rays from extragalactic flares will be challenging to detect, even with next generation gamma-ray wide-field observatories. 

The delayed gamma-ray emission component from SGR 1806-20 was not sufficiently bright, given the limited sensitivity and lack of spectral information of the {\it Konus-WIND} or {\it RHESSI} observations, to search for individual decay line features in the energy spectrum, the ``smoking gun'' for $r$-process production (Fig.~\ref{fig:spectra}). Prospects may be better for the next Galactic giant flare, depending on the landscape of the next generation satellites. The Compton Spectrometer and Imager (COSI; \citealt{Tomsick+23}), with an anticipated launch date of 2027, will have sensitivity in the 0.2-5 MeV energy range to gamma-ray lines with an energy resolution $\Delta E/E \lesssim 1\%$. For a comparable gamma-ray signature as that seen following the SGR 1806-20 giant flare, COSI would have a strong detection of the signal, allowing for spectral characterization and searches for individual bumps at early times. It may be able to detect the signal until $\sim$30~ks after the onset of the giant flare. 

Although the $r$-process decay spectrum peaks at $\sim$ MeV energies, several decay lines occur at hard X-ray energies $\sim 10$ keV \citep{Ripley+14}, albeit carrying a luminosity roughly 2-3 orders of magnitude smaller than the total.  Furthermore, due to the much higher bound-free opacity $\kappa_{\rm X} \gtrsim 10^{2}$ g cm$^{-2}$ of $r$-process nuclei at $\sim 10$ keV \citep{Barnes+16}, the X-ray light-curve would not peak until $t_{\rm X} \sim (\kappa_{\rm X}/\kappa_{\gamma})^{1/2}t_{\rm peak} \gtrsim 3\times 10^{4}$ s, when the decay luminosity is smaller by a factor of $\gtrsim 100$ than at $t_{\rm peak}$, such that $L_{\rm X}(t_{\rm X}) \sim 10^{34}-10^{35}$ erg s$^{-1}$ for $M_{\rm r} \sim 10^{-6}M_{\odot}$. Nevertheless, given the greater sensitivity of hard X-ray telescopes such as NuSTAR \citep{Harrison+13_nustar} than those in the MeV window, an X-ray detection of an $r$-process decay signal might be possible for a future SGR 1806-20-like event. 

There are next-generation mission concepts in the MeV range which are significantly more sensitive than COSI, with comparable energy resolution (e.g., \citealt{Timmes+19}). There are two possible methods. One is the adaptation of liquid argon technology, utilized for direct dark matter experiments, for use in space (e.g., \citealt{Aramaki+20}). The other concept are narrow-field nuclear spectro(polari)meters, utilizing techniques like multilayer optics or Laue lenses. It may be feasible for such missions to detect the $r$-process signature of giant flares for months or longer after the event.

\subsection{As Sources of $r$-Process Cosmic Rays}

As the ejecta collides and shocks the gaseous medium surrounding the magnetar, electrons are accelerated into a non-thermal (e.g., power-law) energy distribution that powers the synchrotron radio afterglow \citep{Gelfand+2005,Granot+2006}. At the same time, the strong reverse shock driven back through the ejecta, accelerates a fraction of the $r$-process nuclei into a similar power-law distribution extending to relativistic energies. Our results thus implicate magnetars as novel sources of $r$-process cosmic rays.  

Although the kinetic energy of the giant flare ejecta $\sim 10^{46}$ erg is $\sim 5$ orders of magnitude smaller than a supernova $E_{\rm SN} \sim 10^{51}$ erg, its mass is $\sim 10^{7}$ times smaller $(M_{\rm ej} \sim 10^{-6}M_{\odot}$ vs. $M_{\rm SN} \sim 10M_{\odot})$, rendering the mean energy per particle $\sim 100$ times higher than in supernovae. Furthermore, while the magnetar ejecta is predominantly composed of $r$-process nuclei, the $r$-process mass-fraction of the matter encountered by supernova shocks is only $X_{\rm r} \sim 10^{-5}$--$10^{-7}$ (if supernovae are themselves $r$-process sources; e.g., \citealt{Komiya+17}). Given also their comparable rates, magnetar giant flares could well dominate supernovae as heavy cosmic ray sources. Neutron-star mergers also generate nearly pure $r$-process ejecta at speeds $v \gtrsim 0.1$--$0.3c$, and their total contribution to Galactic $r$-process production likely eclipses magnetars (Sec.~\ref{sec:chem_evolution}).  However, because mergers are far rarer, occurring once every $\sim 10^{4}$--$10^{5}$ yr in the Milky Way (e.g., \citealt{Kalogera+04}), their cosmic ray flux at Earth fluctuates by orders of magnitude over timescales of a few Myr, due to the short confinement time of cosmic rays in the Galactic plane relative to the local merger rate, spending most time at a level too low to account for observations (e.g., \citealt{Komiya+17}). Taken together, our findings support magnetars as major if not dominant contributors to the $r$-process cosmic ray flux at Earth.

\subsection{Alternative Models: Non-thermal Emission?}

With the current data, it is challenging to definitively exclude other interpretations for the delayed MeV component from SGR 1806-20, such as emission from non-thermal particles.  For example, a hard spectrum peaking at MeV energies is, at first glance, similar to the phenomenological ``Band'' function used to characterize gamma-ray burst (GRB) spectra \citep{Band+93}. Importantly, however, the timescale, luminosity and variability properties of the late-time SGR 1806-20 emission are qualitatively different from GRBs. 

%Similarly, one might consider a non-thermal origin for this emission component from shock interactions with a circumstellar medium (CSM). However, explaining the observed rise-time as the deceleration time of the baryon ejecta shell $M_{\rm ej} \gtrsim 10^{25}$ g \citep{Gelfand+2005} would demand a CSM mass $\gtrsim M_{\rm ej}$ that is located $\lesssim t_{\rm peak}v_{\rm ej} \sim 10^{13}$\,cm from the magnetar prior to the giant flare. Further, if this material was in the form of a steady wind, it would require an enormous mass-loss rate, $\dot{M} \sim M_{\rm ej}/t_{\rm peak} \sim 10^{-4}M_{\odot}$ yr$^{-1}$). 

Resonant scattering by outflowing particles from the magnetar magnetosphere can create hard gamma-ray emission extending up to MeV energies \citep{Beloborodov13}. However, the expected luminosities are $3$--$4$ orders of magnitude below that observed from SGR 1806-20. The emission would also be expected to be strongly anisotropic and hence modulated on the rotation period of the neutron star; yet, such a periodicity was searched for and not observed in the SGR 1806-20 late-time MeV emission \citep{Boggs+07}. Thus, while non-thermal models cannot be definitively excluded, the model of $r$-process decay forwarded here provides a consistent explanation for the timescale, luminosity, decay-rate, and spectrum, and it makes direct predictions for future observations.

The source was spatially proximal with the Sun at the time of the GF, and \citet{Boggs+07} noted a change in flux around the time of sunset, raising the possibility of a solar origin for the delayed MeV signal. However, as also noted by \citet{Boggs+07}, the signal hardness and duration would be completely novel for a solar flare (where the hardest signals typically peak below a few hundred keV and the longest signals last under half an hour; e.g., \citealt{ursi2023first,knuth2020subsecond}).
We estimate that the probability for such a never-before-seen solar flare to be coincident within $\sim 10^{3}$ s of a rare celestial event like a magnetar GF to be $\lesssim 5\times 10^{-7}$ (corresponding to $\sim5\sigma).$ Furthermore, in Appendix \ref{app:solar}, we present GOES X-ray data indicating no abnormal solar activity took place during the time of the GF.

\section{Summary}

The previously unexplained delayed MeV emission component from the SGR 1806-20 giant flare \citep{Mereghetti+05,Frederiks+07,Boggs+07} is naturally explained as gamma-ray line emission from freshly synthesized radioactive $r$-process nuclei expelled into space during the flare. The rise-time, decay rate, and fluence of the light-curve (Fig.~\ref{fig:lightcurve}), as well as the hard energy spectrum peaking around $\sim$ 1 MeV (Fig.~\ref{fig:spectra}), all agree with those predicted from radioactive gamma-ray decay lines for $\sim10^{-6}M_{\odot}$ of $r$-process ejecta. 

The finding that magnetars produce heavy elements, as just the second directly confirmed $r$-process source after neutron star mergers, has implications for the chemical evolution of the Galaxy. In particular giant flares offer a confirmed source that promptly tracks star-formation. The light curve and continuum spectral shape of the SGR 1806-20 gamma-ray emission are alone insufficient to discern details of the synthesized abundances, such as the relative mass fractions of the three $r$-process peaks. These are sensitive to the details of the ejection mechanism and the electron fraction of the ejecta, which depends on the depth and covering fraction of the mass excavated from the neutron star crust (\citetalias{Cehula+24}, their Fig.~1). The opportunity to directly study the earliest stages of $r$-process nucleosynthesis at the isotope-by-isotope level further motivates the development and deployment of MeV-sensitive instruments \citep{Timmes+19} in preparation for the next Galactic giant flare.

A straightforward prediction of the model is kilonova-like UV/optical emission ({\it nova brevis}) with a timescale of minutes and peak luminosity $\lesssim 10^{40}$ erg s$^{-1}$, similar to that of the gamma-ray emission (\citetalias{Cehula+24}; \citealt{Patel+2025}). Specifically, for the fiducial ejecta model which matches the SGR 1806-20 gamma-ray emission (Fig.~\ref{fig:lightcurve}), we predict a peak g-band magnitude $m_{\rm AB} = 4$ at 15 minutes.  While this is exceptionally bright and in principle visible to the naked-eye, this signal could not be seen for SGR 1806-20 because of its location close to the Sun at the time of the flare and large line of site extinction. Despite their rapid evolution, {\it novae breves} may be detected by either monitoring Galactic magnetars continuously from the ground, or by rapidly slewing to the next giant flare in response to the prompt spike gamma-ray trigger. 

Several multimessenger signals are also predicted. Radioactive decay of $r$-process nuclei also powers an $\sim$ MeV neutrino signal which peaks roughly a second after the flare, though at a luminosity too low to detect with existing or planned detectors. A more promising neutrino source is the hot crust of the neutron star shocked during the flare (\citetalias{Cehula+24}). Asymmetric ejection of a significant quantity of mass should also produce a gravitational wave signal which may be detectable by future ground-based observatories (e.g., \citealt{Beniamini+2024}).

The gamma-ray signal described here was predicted during the course of our explorations of $r$-process nucleosynthesis in magnetar giant flares (\citetalias{Cehula+24}; \citealt{Patel+2025}). We came to recognize its presence in the SGR 1806-20 flare only after ``rediscovering'' the (previously unexplained and largely forgotten) delayed gamma-ray component in the literature. In retrospect, however, heavy-element nucleosynthesis is an inevitable consequence of sudden mass ejection from a neutron star (e.g., \citealt{Lattimer+1977}), for which there was already strong evidence from the giant flare radio afterglows (e.g., \citealt{Frail+1999,Gelfand+2005}). Thus, in the likely case that our understanding of the process of baryon ejection in magnetar giant flares changes over time, the interpretation drawn here is likely to remain robust. At a basic level, the gamma-ray signal depends only on the ejecta velocity distribution, and the fact that such a large mass $\gtrsim 10^{-7}$--$10^{-6}M_{\odot}$ must necessarily be excavated from depths below the neutron star surface which are sufficiently neutron-rich to enable the synthesis of radioactive nuclei.

\vspace{12pt}
% \begin{acknowledgments}
We thank Yong Qian for encouragement and helpful conversations. A.~P.~and B.~D.~M.~were supported in part by the National Science Foundation (AST-2009255) and by the NASA Fermi Guest Investigator Program (80NSSC22K1574). The Flatiron Institute is supported by the Simons Foundation. This research was supported in part by grant no. NSF PHY-2309135 to the Kavli Institute for Theoretical Physics (KITP). J.~C. was supported by the Charles University Grant Agency (GA UK) project No. 81224. This work was initiated in part at the Aspen Center for Physics, which is supported by a grant from the Simons Foundation (1161654, Troyer).
% \end{acknowledgments}

\appendix

\section{Reanalysis of Gamma-Ray Data}
\label{sec:app}

The discovery of this signal was first reported in analysis of {\it INTEGRAL} SPI-ACS data \citep{Mereghetti+05}. This component was given the name ``afterglow'' due to the associated radio signature identified in this event and also the giant flare from SGR 1900+14. The signal was subsequently identified in data from {\it Konus-WIND} \citep{Frederiks+07} and the back detectors on {\it RHESSI} \citep{boggs2007giant}. Because this signature was discovered without the context of $r$-process events, the originating papers did not perform the key analysis for this possibility, i.e., quantifying the temporal decay of the signal past peak and inference on the total fluence. We thus reanalyze the reported information in this context, utilizing information from all three detections.

\citet{Mereghetti+05} identified the signal as becoming dominant from $t = 400\,\rm s$ and lasting until $t \approx 4000\,\rm s$. The power law fit for the flux reported ($\propto t^{-\delta},$ $\delta = 0.85$) was conducted over the whole ``MeV afterglow" phase, starting from $t = 500\,\rm s$, prior to peak. Insofar as this does not isolate the decay phase and thus does not directly probe the radioactive decay rate, we repeat the fit starting from peak, between $t \approx 675$--$12000\, \rm s$. This analysis is performed on background-subtracted data, in order to isolate temporal evolution of the signal by removing the temporal evolution of the background. We fit a linear background from $t_0 - 2000\,\rm s$ to $t_0 -200\,\rm s$ and from $t_0 +12000\,\rm s$ to $t_0 + 14000\,\rm s$, given the identification of the signal out to $t = 12000\,\rm s$ in {\it Konus-WIND} \citep{Frederiks+07}. 

\begin{figure}
   \centering
   \includegraphics[width=1.0\textwidth]{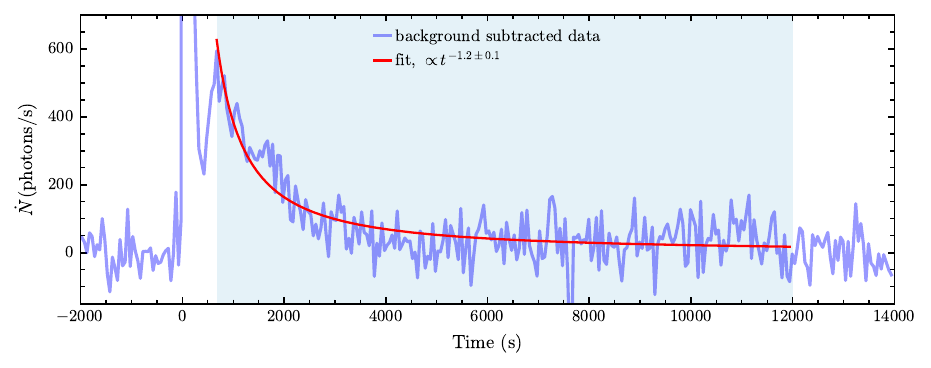}
   \caption{SPI-ACS decay fit from 675 s (peak) to 12000 s.}
   \label{fig:fit}
\end{figure}

Based on this analysis, we report a measurement of the power law index in the decay phase, from $t_0 +675\,\rm s$ to $t_0 +12000\,\rm s$, of $\delta = 1.2 \pm0.1$ (1$\sigma$), as shown in Fig.~\ref{fig:fit}. A similar value is obtained when fitting the decay to the original end of $t_0 +4000\,\rm s$ utilized in the SPI-ACS discovery paper.  This power-law index is within the range for that expected for $r$-process decay, $\alpha = 1.1$--$1.4$ based on previous studies (e.g., \citealt{Metzger+10}), and to the decay rate found in our simulations of the MeV light-curve, $\alpha \approx 1.2$ (Eq.~\eqref{eq:analytic_qdot}).

The other key parameter of interest to interpret the signal is the brightness. {\it INTEGRAL} did not quantify fluence due to the lack of spectral information in SPI-ACS and had only partial coverage \citep{Mereghetti+05}. {\it Konus} provided a fluence estimate at only late times due to a data gap, and assumed a spectrum given the limited spectral channels available at this time \citep{Frederiks+07}. {\it RHESSI} fit an empirical spectral model but has coverage of only the first few hundred seconds \citep{boggs2007giant}. Thus, we must infer the total brightness by combining information from the separate detections.

Typically, this cannot be done with reasonable accuracy because the response matrices are not-invertible, requiring a forward-folding approach which assumes a spectral model. However, we benefit from the lack of strong spectral evolution from the expected nucleosynthetic signature over the time range of interest. Thus, we can utilize the {\it RHESSI} spectral measurement and the {\it INTEGRAL} SPI-ACS count rate information in the same interval to create a brightness-to-counts mapping in SPI-ACS. We determine the ratio of the number of counts in SPI-ACS during this interval over the number of counts in SPI-ACS in the total interval. Then, the total fluence is the {\it RHESSI} reported fluence over the sub-interval scaled by this ratio. 

Specifically, {\it RHESSI} reports a fit to thermal bremsstrahlung with $k_{\rm B}T\approx$1.9~MeV \citep{boggs2007giant}, with a fluence of ($1.85\pm0.25$)$\times 10^{-4}~\rm erg\, cm^{-2}$ as measured over the $t_0+400\,\rm s$ to $t_0+900\,\rm s$ interval. The counts in SPI-ACS during this interval are $\sim$21\% of the total counts over the $t_0+400\,\rm s$ to $t_0+12000\,\rm s$ interval. Thus, we infer a total fluence of ($8.7\pm1.2$)$\times 10^{-4}~\rm erg\, cm^{-2}$. Given the approximate calculation method and lack of consideration for intercalibration uncertainty it is reasonable to assume the systematic error is a similar size as the statistical error. As a sanity check, we repeat the procedure to predict the fluence seen in the $t_0+5000\,\rm s$ to $t_0+12000\,\rm s$ interval observed by {\it Konus}, which gives a value in agreement with their reported fluence.

\section{Association of the signature to SGR 1806-20}\label{app:solar}

\begin{figure}
   \centering
   \includegraphics[width=1.0\textwidth]{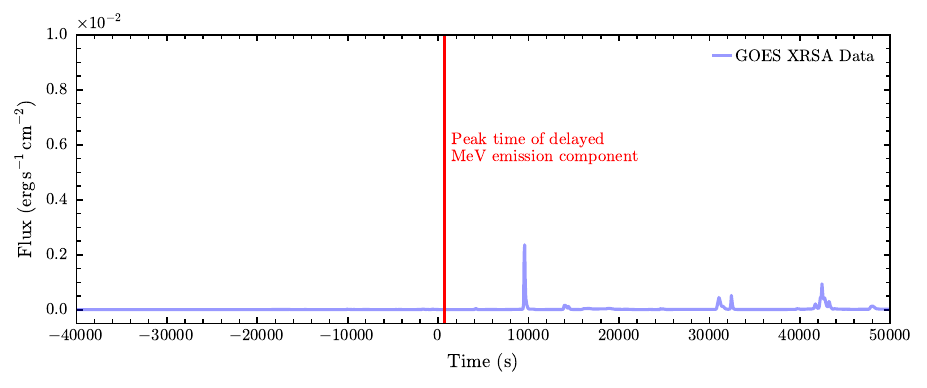}
   \caption{GOES X-ray data (3-25~keV) around the GF time. The solar X-ray flux is classified at background activity levels during the time of the delayed MeV emission; there is no solar activity which could explain the putative $r$-process signal.}
   \label{fig:goes}
\end{figure}

As the detecting instruments of the gamma-ray signal do not have precise localization ability, it is, in principle, possible that the gamma-ray flare which is evident from $t_0+400~\rm s$ to $t_0+120000~\rm s$ could arise from an unrelated source, occurring in temporal proximity by chance. As noted, \citet{boggs2007giant} mention the Sun was near the position of SGR 1806-20 at the time of the GF. In order to explore a solar origin, we gather X-ray data from the GOES-12 satellite \citep{hill2005noaa} for a day around the GF trigger time, as shown in Figure~\ref{fig:goes}. A gamma-ray solar signature would have a corresponding X-ray signature, owing both to the intrinsic photon spectrum as well as the relative sensitivities of the narrow-field X-ray telescope as compared with the all-sky gamma-ray monitors. The lack of contemporaneous X-ray emission thus fully excludes a solar origin for this event.

This is not surprising because the temporal and spectral characteristics of the signature explored here is incompatible with known solar signatures. Regular solar-flare activity are regularly seen in gamma-ray burst monitors. These last for $\sim$10~minutes and have peak energies below the low-energy threshold of these instruments \citep[e.g.][]{ursi2023first}. A separate signature is observed from reconnection events. These have harder spectra, up to a few hundred keV, but last under a few seconds in duration \citep[e.g.][]{knuth2020subsecond}. Thus, were the putative $r$-process signature to be solar, it would require an event which does not emit also in X-rays, and which has not been identified in the 60~years of continuous coverage of the gamma-ray sky \citep{burns2023gamma}. The chance of such a signal occurring within $\sim 10^3$~s of the GF trigger time is $\sim 10^3$~s$/60$~years $\approx 5\times 10^{-7}$.

We lastly explore any other possible explanation from known signatures. Given the detection in multiple spacecraft, the origin cannot be due to local particle activity nor terrestrial emission. The spectral hardness precludes any known Galactic signature. No gamma-ray burst identified has a comparable duration and peak energy. The unusual properties of this signature, incompatibility with known gamma-ray transient properties, and the likelihood of seeing a signature only once in decades of observation are why the original discovery and characterization papers only minimally explore this question. Thus, the calculation above captures the chance coincidence of observing any signature unrelated to SGR 1806-20, and we are confident in the association.

\section{Parameter Exploration}
\label{app:param}

\begin{figure}[t]
   \centering
   \includegraphics[width=0.993\textwidth]{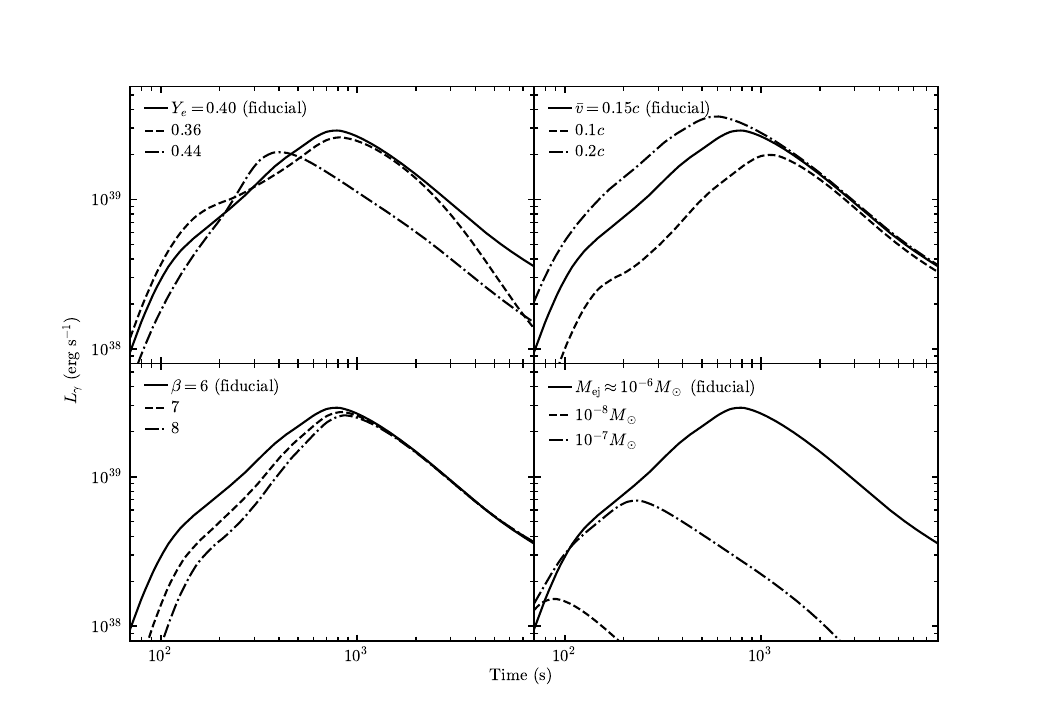}
   \caption{Parameter study exploring the isolated affects of four key parameters: $Y_e$ (top left), $\bar{v}$ (top right), $\beta$ (bottom left), $M_{\rm ej}$ (bottom right). The fiducial model is shown for comparison with a solid black line ($M_{\rm ej} \approx 1.2 \times 10^{-6},\, Y_e = 0.40, \, \bar{v} = 0.15,\,\beta = 6$). For each model, the parameter labeled in the legend is the only one that is changed; all other parameters are held fixed at their fiducial value.}
   \label{fig:params}
\end{figure}

Fig.~\ref{fig:params} shows the effect of varying the key model parameters entering our nucleosynthesis and light-curve calculations. Each panel shows a model in which a single parameter $\{M_{\rm ej}, Y_e, \bar{v}, \beta\}$ has been varied from its value in the fiducial model ($M_{\rm ej} \approx 1.2 \times 10^{-6},\, Y_e = 0.40, \, \bar{v} = 0.15,\,\beta = 6$), which we show for comparison with solid black lines.

The electron fraction of the ejecta depends on the depth below the neutron star surface excavated in mass ejected during the flare and any weak interactions that occur during the ejection process. We find the effects of changing $Y_e$ (top left panel) may affect both the peak time-scale and decay rate of the MeV emission. In particular, higher $Y_e > 0.40$ results in a lower fraction of the total ejecta mass being synthesized into $r$-process nuclei, thus reducing the radioactive gamma-ray emission. A lower electron fraction $Y_e < 0.40$ changes the distribution of mass between the $r$-process abundance peaks by producing more heavy neutron rich nuclei and fewer light seed nuclei, therefore steepening the late-time decay rate.

The bulk velocity $\bar{v}$ (top right panel) controls the early time magnitude and peak timescale of the transient. Faster expanding ejecta ($\bar{v} = 0.2c$) decompresses more quickly and become optically-thin earlier when the energy generation rate from decaying nuclei is larger (Eq.~\eqref{eq:analytic_qdot}), resulting in earlier and brighter peak emission (Eq.~\eqref{eq:tthin}). The opposite is true for slower ejecta ($\bar{v} = 0.1c$).

The parameter $\beta$ (bottom left panel) dictating the ejecta velocity distribution (Eq.~\eqref{eq:rho}) primarily affects the shape of the early optically-thick phase of the light-curve. For larger values of $\beta$, less mass is concentrated in the fastest outer ejecta layers and is instead shifted to deeper slower ejecta. The outer layers, which become optically-thin first, contain less radioactive material and the photosphere (Eq.~\eqref{eq:R}) will encounter more mass as it recedes to the inner layers, resulting in a steeper light-curve rise and slightly delayed and dimmer peak.

Finally, we show models with lower total ejecta mass $M_{\rm ej} \gtrsim 10^{-8}M_\odot$ (bottom right panel), spanning the total range found by modeling the radio afterglow of the SGR 1806-20 giant flare. Lower ejecta mass significantly reduces the luminosity of the $r$-process gamma-ray emission (Eq.~\eqref{eq:analytic_Qgamma}), indicating that an ejecta mass close to the fiducial value $\sim 10^{-6}M_\odot$ is required to explain the delayed MeV emission.

\section{Partition of $\beta$-Decay Energy into Gamma-Rays}
\label{sec:partition}
\begin{figure}[htbp]
   \centering
   \includegraphics[width=0.5\textwidth]{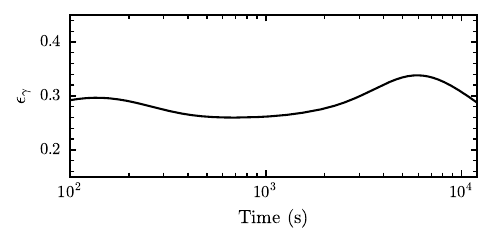}
   \caption{The fraction $\epsilon_\gamma$ of $\beta$-decay energy emitted as $\gamma$-rays for the fiducial model during over the time-range of interest (Fig.~\ref{fig:lightcurve}).}
   \label{fig:epsilon}
\end{figure}

In kilonova outflows (characterized by $Y_e \lesssim 0.25$), a fraction $\epsilon_\gamma \approx 0.4$ of the $\beta$-decay energy partitions into $\gamma$-rays (with the remaining energy in electrons and neutrinos) on timescales $0.1$--$1\,\rm days$ (\citealt{Barnes+16}, see their Fig.~4). We conduct the analysis (Sec.~\ref{sec:spectrum}) on the timescale of $\sim$ minutes and for the higher electron fractions of relevance to the delayed MeV emission studied in this work. We show $\epsilon_\gamma(t) = \dot{q}_\gamma(t)/\dot{q}_{\rm r}(t)\approx 0.3$ in Fig.~\ref{fig:epsilon}, where $\dot{q}_\gamma$ is calculated via Eq.~\eqref{eq:gamma}. Given our model $Y_e = 0.40$, this value trends consistently with results from \citet{Barnes+16} where $\epsilon_\gamma$ was shown to decrease with rising electron fraction.

\bibliographystyle{aasjournal}
\bibliography{refs,example,refs2}

\end{document}